\documentclass[sn-mathphys,Numbered]{sn-jnl}

\usepackage{graphicx}%
\usepackage{multirow}%
\usepackage{amsmath,amssymb,amsfonts}%
\usepackage{amsthm}%
\usepackage{mathrsfs}%
\usepackage[title]{appendix}%
\usepackage{xcolor}%
\usepackage{textcomp}%
\usepackage{manyfoot}%
\usepackage{booktabs}%
\usepackage{algorithm}%
\usepackage{algorithmicx}%
\usepackage{algpseudocode}%
\usepackage{listings}%

\theoremstyle{thmstyleone}%
\theoremstyle{thmstyletwo}%

\theoremstyle{thmstylethree}%

\begin{document}

\title[Effect of Rydberg-atom-based sensor performance on different Rydberg atom population at one atomic-vapor cell]{Effect of Rydberg-atom-based sensor performance on different Rydberg atom population at one atomic-vapor cell}



\author[1]{\fnm{Bo} \sur{Wu}}

\author*[1]{\fnm{Qiang} \sur{An}}\email{anqiang18@nudt.edu.cn}
\equalcont{These authors contributed equally to this work.}

\author[2]{\fnm{Jiawei} \sur{Yao}}

\author[1]{\fnm{Fengchuan} \sur{Wu}}

\author[1]{\fnm{Yunqi} \sur{Fu}}

\affil*[1]{\orgdiv{College of Electronic Science and Technology}, \orgname{National University of Defense Technology}, \orgaddress{\city{Changsha}, \postcode{410073},  \country{China}}}

\affil[2]{\orgdiv{Northwest Institute of Nuclear Technology},  \orgaddress{\city{28 Pingyu Road Xi'an}, \postcode{710024}, \state{Shaanxi}, \country{China}}}


\abstract{The atomic-vapor cell is a vital component for Rydberg atomic microwave sensors, and impacts on overall capability of Rydberg sensor. However, the conventional analysis approach on effect of vapor-cell length contains two implicit assumptions, that is, the same atomic population density and buffer gas pressure, which make it unable to accurately capture actual response about effect of Rydberg-atom-based sensor performance on different Rydberg atom population. Here, utilizing a stepped cesium atomic-vapor cell with five different dimensions at the same atomic population density and buffer gas pressure, the height and full width at half maximum of Electromagnetically Induced Transparency(EIT) signal, and the sensitivity of the atomic superheterodyne sensor are comprehensively investigated at the same Rabi frequences(saturated laser power) conditions. It is identified that EIT signal height is proportional to the cell length, full width at half maximum and sensitivity grow with the increment of cell length to a certain extent. Based on the coherent integration signal theory and atomic linear expansion coefficient method, theoretical analysis of the EIT height and sensitivity are further investigated. The results could shed new light on the understanding and design of ultrahigh-sensitivity Rydberg atomic microwave sensors and find promising applications in quantum measurement, communication, and imaging.}




\maketitle

\section{Introduction}\label{sec1}
\section{Introduction}

In the past ten years, Rydberg atomic microwave sensor\cite{gordon2014millimeter,holloway2014broadband,paradis2019atomic} has made significant progress due to ravishing characteristics that classical radio frequency(RF) microwave(MW) technology does not have, including self-calibration\cite{holloway2014sub}, high sensitivity\cite{anderson2017continuous,fan2014subwavelength}, broad operation frequency\cite{holloway2014broadband}. It has a wide range of applications, such as sensors for communication signals (amplitude modulated\cite{sedlacek2013atom,simons2021continuous2}, frequency modulated\cite{robinson2021determining}, phase modulated signals\cite{anderson2021self}), Rydberg Microwave Frequency Comb spectrometer\cite{Microwave}, imaging \cite{Berchera}, quantum secured communications\cite{CHUNNILALL}and so on.

It has been theoretically demonstrated that measurement sensitivity limit of Rydberg atomic microwave sensor is -220 dBm/Hz \cite{fan2015atom}, which is far exceeding classical sensor sensitivity limit of -174 dBm/Hz (room temperature)\cite{holloway2019detecting}. Further studies have put forward solutions to improve experimental measurement sensitivity including atomic superheterodyne sensor\cite{jing2020atomic}, enhanced metrology at the critical point of a many-body Rydberg atomic system\cite{Dong-Sheng}, resonator\cite{holloway2022rydberg, wu2022design}, and multi-photon readout\cite{HQ}. According to the quantum projection noise-limited sensitivity formula\cite{1999Quantum}, sensitivity grows by increasing the Rydberg atom population.  Due to finite Alkali metal atoms density, longer cell length is often utilized to achieve larger Rydberg atom population. But there remains elusive in no good consistency for multiple different length cells. Essentially, multiple different length cells’ atomic population density and pressure of buffer gas is differential which make it unable to accurately capture actual response about Rydberg atom population effect.

In this paper, we propose a stepped atomic-vapor cell with five different path lengths under the same atomic population density, buffer gas pressure. We rigorously study how the Rydberg atom population affect the Electromagnetically Induced Transparency (EIT) signal height, full width at half-maximum(FWHM) and sensitivity of superheterodyne atomic sensor, respectively in saturated laser power conditions. First of all, when coupling, probe laser power reaches a specific value, the EIT height reaches saturation to investigate effect of Rydberg atom population. Additionally, we theoretically and comprehensively analyze effect of different Rydberg atom population by utilizing coherent signal integration theory and atomic linear expansion coefficient method. The experimental and theoretical results verify the EIT signal height, FWHM, and sensitivity increase with Rydberg atom population in the case of shorter length cell. Specifically, the EIT signal height keeps a steady proportional upward trend with the cell length from 5 to 25 mm. Our results may provide helpful suggestions for the high-sensitivity Rydberg atomic microwave sensor design for practical applications.

\section{Experimental setup}
\begin{figure}[!h]\centering
	\includegraphics[width=0.5\textwidth]{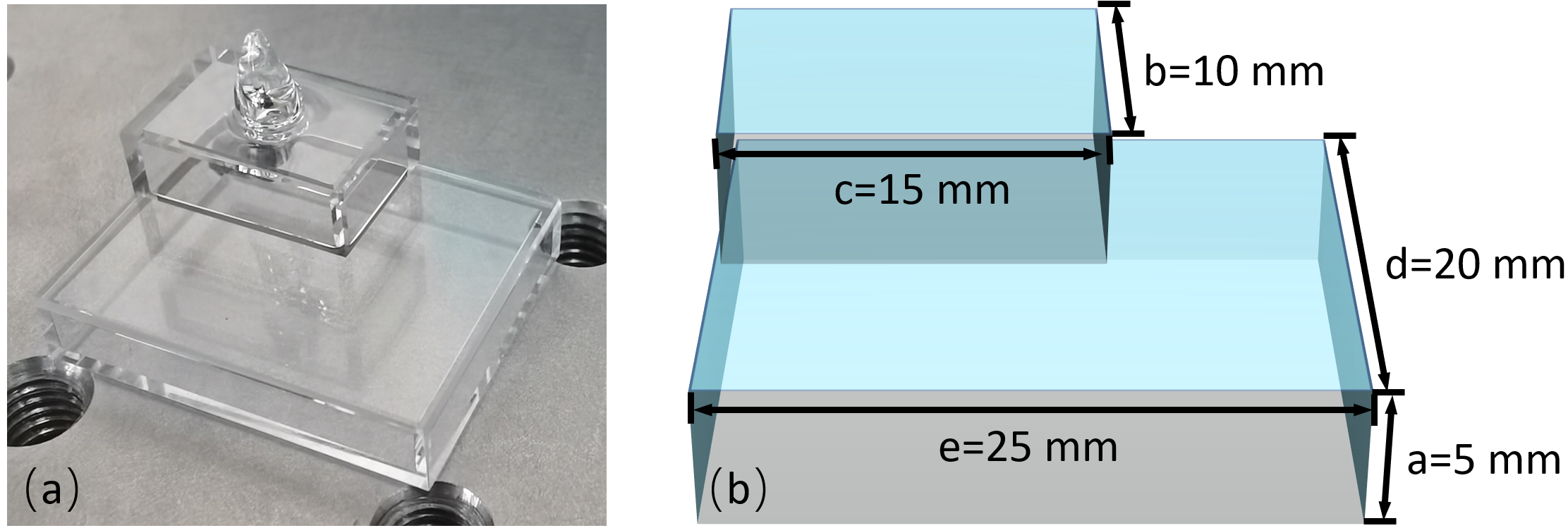}
	\caption{\label{1}  Geometrical details of the stepped atomic-vapor cell are presented. (a)Measured prototype from front view. (b) Specific size of stepped atomic-vapor cell(a=5 mm, b=10 mm, c=15 mm, d=20 mm,e=25 mm.) }
\end{figure}

As shown in Figure \ref{1}(a), the stepped atomic-vapor cell structure is adopted to investigate the Rydberg state population effect on Rydberg atomic microwave sensors. The stepped atomic-vapor cell is composed of Schott BOROFLOAT material and has five different dimensions under the same atomic population density and buffer gas pressure. Figure \ref{1}(a) presents measured prototype geometry. To clarify structure of the stepped atomic-vapor cell more clearly, specific size is illustrated in Figure \ref{1}(b). 
\begin{figure}[!h]\centering
	\includegraphics[width=0.8\textwidth]{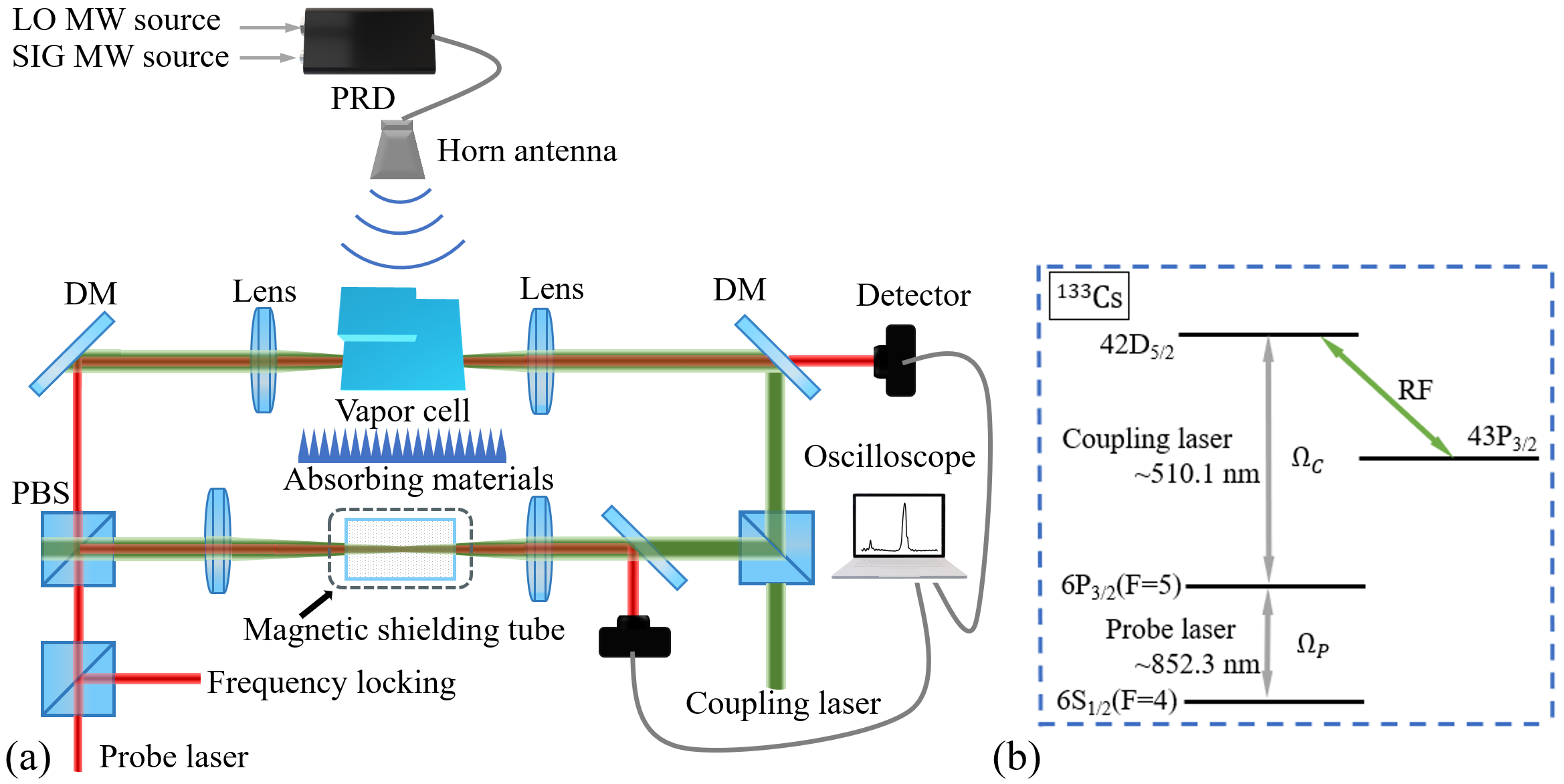}
	\caption{\label{fig2}(a) Overview of experimental setup. We have also utilized following notations:(1) DM:dichroic mirror, (2) PBS:polarizing beam splitter, (3) RPD:2-way MW resistive power divider, (4) LO MW source:local oscillator microwave source, and(5) SIG MW source:weak signal microwave source. (b)Overview of experimental energy diagram. An 852.3 nm probe laser excites cesium atoms from ground state $6\textup{S}_{1/2}$(F=4) to intermediate state $6\textup{P}_{3/2}$(F=5), a 510.1 nm coupling laser drives atoms from intermediate state to the Rydberg state $42\textup{D}_{5/2}$, and the 9.93 GHz radio frequency(RF) MW stimulates atoms from Rydberg state $42\textup{D}_{5/2}$  to Rydberg state $43\textup{P}_{3/2}$ }
\end{figure}

As shown in Figure \ref{fig2}(b), cesium atoms can be easily excited from ground to Rydberg states utilizing two-photon resonance. The experimental setup utilized to measure the effect of Rydberg-atom-based sensor utilizing stepped atomic-vapor cell is shown in Figure \ref{fig2}(a). In the first stage, probe laser and coupling laser counter-propagate and overlap through the center of cell. In the next step, the detector senses probe laser, which converts optical signal to electrical signal. Ultimately, the EIT spectrum can be obtained by scanning at 510.1 nm coupling laser. The EIT signal within magnetic shielding tube serves as reference signal for our system. To generate MW field, we utilize two signal generators with an output power of up to 25 dBm.

\section{Results and Discussion}
\begin{figure}[h]\centering
	\includegraphics[width=0.9\textwidth]{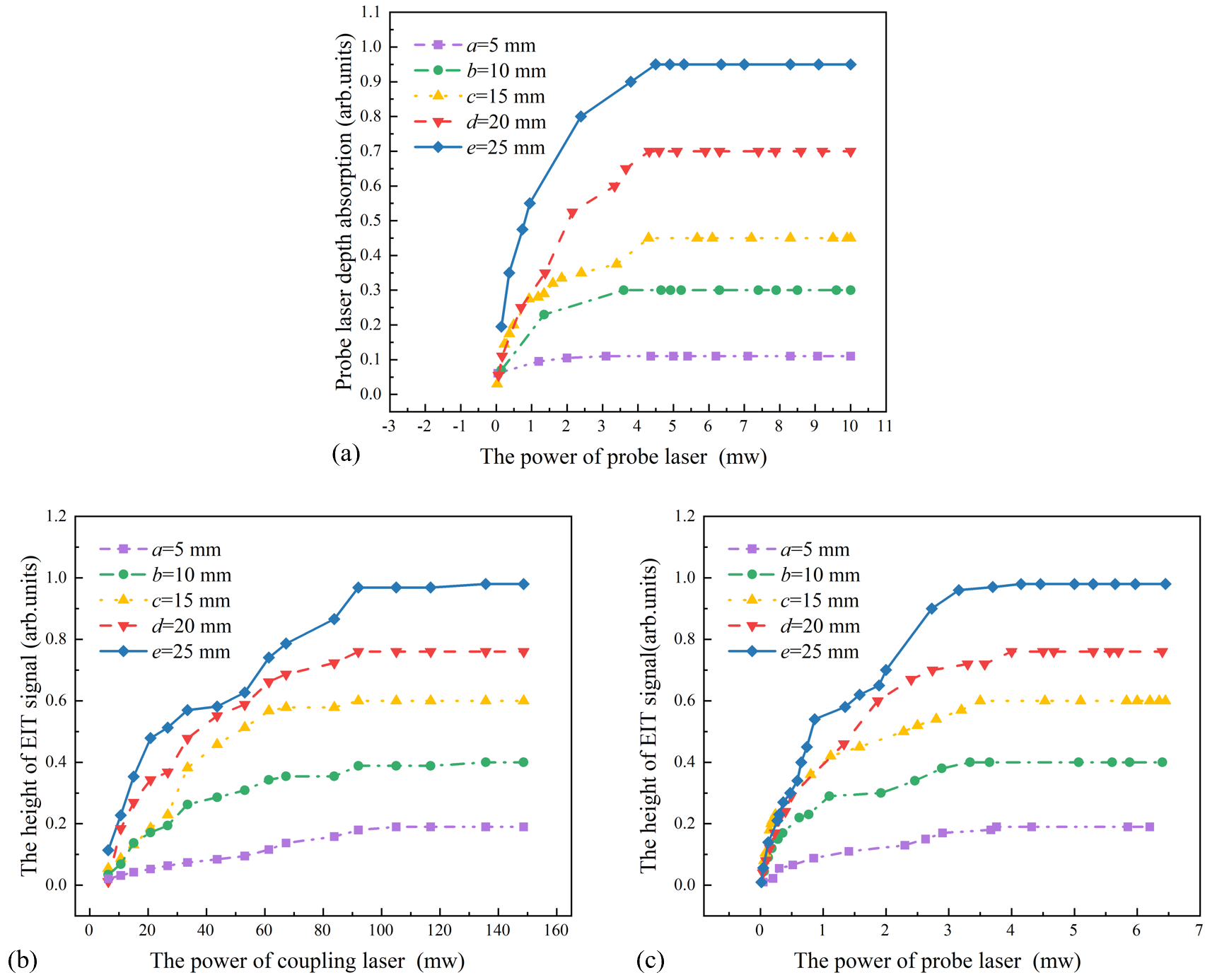}
	\caption{\label{fig3}(a)Depth of probe laser absorption spectrum versus probe laser power.(b)The height of EIT signal versus coupling laser power.(c)The height of EIT signal versus probe laser power. The purple, green, yellow, red and blue curves represent the five lengths of 5 mm, 10 mm, 15 mm, 20 mm and 25 mm respectively in Figure \ref{1}(b)}
\end{figure}

Theoretically, the EIT signal intensity can enhance by increasing laser power, but signal intensity does not grow indefinitely. In order to find a set of saturation coupling and probe laser powers appropriate for five different lengths, Figure \ref{fig3}(a) presents probe laser's power concerning  absorption spectrum's depth. It is seen that absorption depth rises with power, i.e., the more atoms there are, the deeper the absorption at two energy levels. The absorption depth reaches saturation when probe power reaches a specific value. The relationship between coupling, probe laser power, and EIT signal height is investigated as shown in Figure \ref{fig3}(b,c), where the EIT height can be obtained by Lorentzn fitting. It can be seen that the EIT height increases with coupling, and probe laser power, i.e., the more Rydberg atom population and the stronger the EIT signal. When  coupling, probe laser power reaches  specific values which are 105 mW coupling laser driving the ransition above with a 1/e waist of 1200 $\mu$m and 
4.5 mW probe laser driving the ransition above with a 1/e waist of 600 $\mu$m, the EIT height reaches saturation to investigate effect of Rydberg atom population. The resuling coupling, probe laser Rabifrequency(f) are  $\Omega /2\pi $ = 81.01 and 1.69 MHz.
\begin{figure}[h]\centering
	\includegraphics[width=0.9\textwidth]{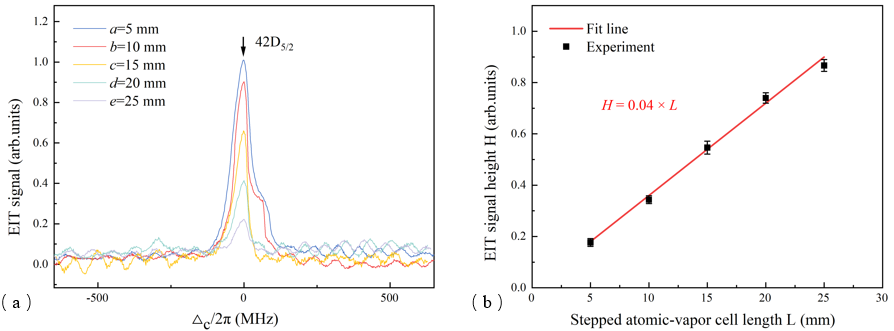}
	\caption{\label{fig4}(a)Measured EIT signals versus coupling laser detuning. (b)Measured EIT signals height versus the cell length. The error bars representing standard deviation of EIT signal height }
\end{figure}

As shown in Figure \ref{fig4}(a), the EIT signals are obtained when coupling, probe laser power is 105 mW, 4.5 mW, respectively. Figure \ref{fig4}(b) presents EIT signals height experience an upward trend from cell length($L$)=5 mm to 25 mm. As expected, the signal is coherent integration while noise is non-coherent. So signal amplitude is proportional to the Rydberg atom population, and noise amplitude is proportional to root of Rydberg atom population. Power is squared to amplitude, so EIT signal height is proportional to Rydberg atom population. Notably, due to the stepped atomic-vapor cell with five different lengths under the same atomic population density, EIT signal height is proportional to the Rydberg atom population involved in the measurement.

Table \ref{table1} summarizes specific evolutions of the EIT full width at half maximum ($\Gamma$) with the increment of cell length. Table \ref{table1} shows that $\Gamma$ is slightly rising with the increment of $L$. The $\Gamma$ is defined by Eq. \ref{eq1}\cite{jing2020atomic}:
\begin{equation}
	\Gamma =\Omega _{p}\sqrt{\frac{2(\Omega _{c}^{2}+\Omega _{p}^{2})}{2\Omega _{p}^{2}+\gamma  ^{2}}}
	\label{eq1}
\end{equation}
\begin{equation}
	\gamma =(\frac{2.405^{2}}{y^{2}}+\frac{\pi ^{2}}{L^{2}})D_{0}\frac{P_{0}}{P}+N_{0}\bar{\nu _{\textup{r}}}\sigma _{1}\frac{P}{P_{0}}+n\bar{\nu _{\textup{Cs}}}\sigma _{\textup{se}}
	\label{eq2}
\end{equation}
where $\Omega _{p}$ is probe Rabi frequency, $\Omega _{c}$ is coupling Rabi frequency and $\gamma$ is longitudinal relaxation rate. Specifically, $\gamma$ can be calculated by Eq. \ref{eq2}\cite{Liu1}, where $y$ is transverse dimensions of the cell, $L$ is cell length, $D_{0}$ is the diffusion constant of cesium in the buffer gas at atmospheric pressure $P_{0}$= 101.3 kPa, $P$ is the buffer gas pressure, $N_{0}$ $\approx $ 2.69×$10^{25}$$ m^{-3}$ is the Loschmidt constant, and $\sigma _{1}$ is the cross section for disorientation. $\bar{\nu _{\textup{r}}}$ is the relative velocity between cesium and buffer gas atoms, $n$ is the cesium populations density which equals $1.62\times 10^{18}/cm^{3}$, $\bar{\nu _{\textup{Cs}}}$ =$\sqrt{(8k_{\textup{B}}T)/(\pi \mu _{\textup{Cs}})}$, T is temperature, $k_{\textup{B}}$ is the Boltzmann constant, $\mu _{\textup{Cs}}$ is the atomic reduced mass and $\sigma _{\textup{se}}$ is the cesium spin-exchange cross-section. Based on the Eqs. \ref{eq1},\ref{eq2}, $\Gamma$ is slightly rising with the increment of $L$.
\begin{table}[h]
	\centering
	\caption{\label{table1}Performance comparisons of $\Gamma$ with the increment of cell length}
	\begin{tabular}{lccccc}
		\toprule  
		& \multicolumn{5}{c}{Experimental} \\   
		\midrule	
		$L$    (mm)      & 5.0      & 10.0    & 15.0  & 20.0  & 25.0 \\
		$\Gamma$ (MHz) & 47.9   & 48.6  & 51  & 56  & 67  \\
		Coupling Rabi f(MHz)  & \multicolumn{5}{c}{81.01}        \\
		Probe Rabi f (MHz)   & \multicolumn{5}{c}{1.69}          \\  
		\bottomrule
	\end{tabular}
\end{table}

\begin{figure}[h]
	\centering
	\includegraphics[width=0.6\textwidth]{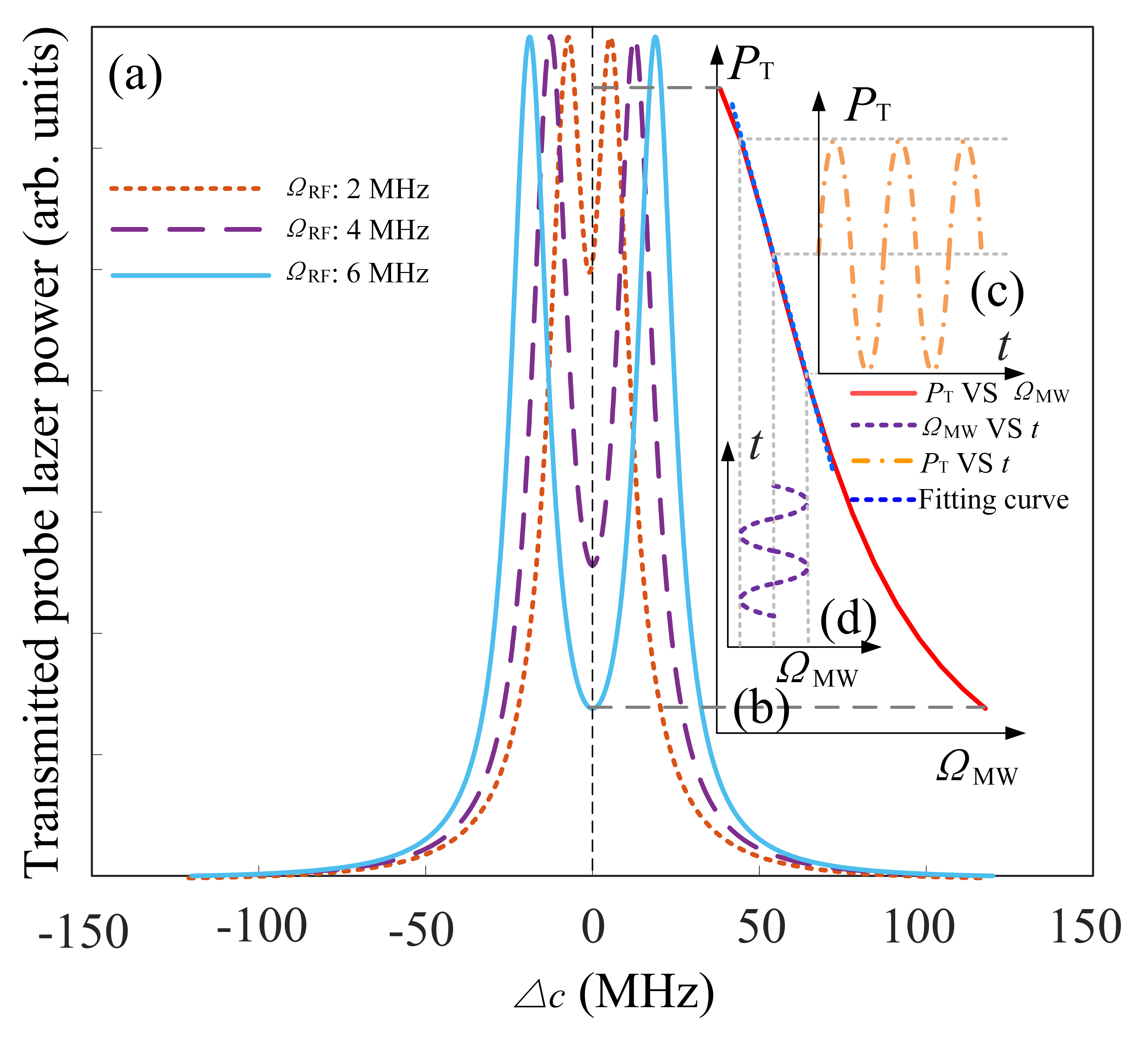}
	\caption{\label{fig555}Numerical simulation response mechanism of Rydberg atom to MW signal }
\end{figure}

In order to investigate the Rydberg atomic microwave superheterodyne sensors sensitivity, which is theoretically defined as minimum detectable power when the signal-to-noise ratio (SNR) decreases to 1, two frequency signal generators are employed as shown in Figure \ref{fig2}(a). The 9.937 GHz resonant frequency is selected as the local oscillator(LO) source, while MW weak signal(SIG) frequency is 9.937 GHz+13 kHz. In the time domain, electric field $E_{\textup{LO}},E_{\textup{SIG}}$ is expressed as:
\begin{equation}
	\left\{\begin{matrix}
		E_{\textup{LO}}=\textup{Acos}(2\pi f_{\textup{LO}}t+\varphi _{\textup{LO}}) \\ 
		E_{\textup{SIG}}=\textup{Bcos}(2\pi f_{\textup{SIG}}t+\varphi _{\textup{SIG}})
	\end{matrix}\right.
	\label{eq333}
\end{equation}

where A and B are the electric field strength of $E_{\textup{LO}}$ and $E_{\textup{SIG}}$, $f_{\textup{LO}}$ and $f_{\textup{SIG}}$ represent the frequency, and $\varphi _{\textup{LO}}$ and $\varphi _{\textup{SIG}}$ are phase, respectively. The 13 kHz detuning frequency means that once the coupling laser is set, the probe laser intensity will oscillate at a beat-note frequency of 13 kHz. The LO and SIG signals are transmitted into a horn antenna through a resistive power divider. The output signal intensity of the detector is measured by a spectrum analyzer with resolution bandwidth of 1 Hz(one-second averaging time). Due to the AC Stark effect, the EIT will have a prominent AT splitting phenomenon when the MW Electric(E)-field is strong. As shown in Figure \ref{fig555}(a), the original EIT peak splits into two transparent absorption peaks, known as two dressed states. The AT splitting is proportional to the E-field and is perceived as\cite{sedlacek2013atom,PhysRevA}:
\begin{equation}
	E=\frac{\hslash}{\mu}\Omega _{\textup{MW}}=2\pi \frac{h}{\mu }\Delta _{\textup{AT}}
	\label{eq44}
\end{equation}
\\where $\Omega _{MW}$ is the Rabi frequency of the Rydberg state transition, $\mu $ is the transition dipole moment of the adjacent Rydberg states, $\hslash$  is reduced Planck’s constant, and $\Delta _{AT}$ is AT split peak spacing. On the basis of the electric field intensity superposition\cite{gordon2019weak}, 
the superposition field $E_{atoms}$ can be expressed as follows:
\begin{align}
	%
	E_{\textup{atoms}}&=E_{\textup{LO}}+E_{\textup{SIG}}\notag\\
	&=\textup{Acos}(2\pi f_{\textup{LO}}t+\varphi _{\textup{LO}})+\textup{Bcos}(2\pi f_{\textup{SIG}}t+\varphi _{\textup{SIG}})
	\label{eq555}
\end{align}
Then, from Euler's formula $e^{j\theta}=\textup{cos}\theta+j\textup{sin}\theta $,
\begin{align}
	%
	E_{\textup{atoms}}&=\textup{Acos}(2\pi f_{\textup{LO}}t+\varphi _{\textup{LO}})+\textup{Bcos}(2\pi f_{\textup{SIG}}t+\varphi _{\textup{SIG}})\notag\\
	&=Re\left\{\textup{A}e^{j(2\pi f_{\textup{LO}}t+\varphi_{\textup{LO}})}+\textup{B}e^{j(2\pi f_{\textup{SIG}}t+\varphi_{\textup{SIG}})}\right\}\notag\\
	&=Re\left\{\textup{A}e^{j(2\pi f_{\textup{LO}}t+\varphi_{\textup{LO}})}\times\left(1+\frac{\textup{A}}{\textup{B}}e^{-j(2\pi\Delta f t+\Delta\varphi)}\right)\right\}
	\label{eq666}
\end{align}

where $\Delta f=f_{\textup{LO}}-f_{\textup{SIG}},\Delta \varphi=\varphi_{\textup{LO}}-\varphi_{\textup{SIG}}$. There is the relationship,

\begin{small}
	\begin{align}
		&\textup{A}\times (1+\frac{\textup{A}}{\textup{B}}e^{-j(2\pi\Delta f t+\Delta\varphi)})	\notag \\  &=\textup{A}+\textup{B}\cos(2\pi\Delta f t+\Delta\varphi)-j \textup{B}\sin(2\pi\Delta f t+\Delta\varphi)\notag \\ 
		&=\sqrt{(\textup{A}+\textup{Bcos}(2\pi\Delta f t+\Delta\varphi))^2+(\textup{B}\sin{(2\pi\Delta f t+\Delta\varphi)})^2}\times e^{j\alpha}
		\label{eq777}
	\end{align}
\end{small}
where $\alpha$ is real number, The $\alpha$ is presented as
\begin{equation}
	|tg\alpha|=\frac{|-\textup{B}\sin{(2\pi\Delta f t+\Delta\varphi)}|}{|\textup{A}+\textup{B}\cos{(2\pi\Delta f t+\Delta\varphi)}| }
	\label{eq787}
\end{equation}

When $\textup{B}\ll \textup{A}$ ,there is $|tg\alpha |\rightarrow 0, \alpha \rightarrow 0$,so it is approximated by Eq.\ref{eq888}

\begin{align}
	&\textup{A}\times (1+\frac{\textup{A}}{\textup{B}}e^{-j(2\pi\Delta f t+\Delta\varphi)})\\ 
	&=\sqrt{(\textup{A}+\textup{B}\cos(2\pi\Delta f t+\Delta\varphi))^2+(\textup{B}\sin{2\pi\Delta f t}+\Delta\varphi)^2}
	\label{eq888}
\end{align}

Put the approximate results Eq.\ref{eq888} back to Eq.\ref{eq666},
\begin{small}
	\begin{align}
		%
		&E_{\textup{atoms}}=Re\left\{\textup{A}e^{j(2\pi f_{\textup{LO}}t+\varphi_{\textup{LO}})}\times\left(1+\frac{\textup{A}}{\textup{B}}e^{-j(2\pi \Delta f t+\Delta\varphi)}\right)\right\}\notag\\
		&=\sqrt{\textup{A}^{2}+\textup{B}^{2}+2\textup{A}\textup{B}\cos(2\pi \Delta f t+\Delta\varphi)}\times\cos(2\pi f_{LO}t+\varphi_{LO})
		\label{eq999}
	\end{align}
\end{small}
When $\textup{B}\ll \textup{A}$, there is $\textup{B}^{2}=\textup{B}^{2}(\textup{cos}(\Delta f t+\Delta \varphi ))^{2}$, the superposition field $E_{atoms}$ can be expressed as follows:
\begin{equation}
	|E_{\textup{atoms}}|\approx A+\textup{Bcos}(2\pi \Delta ft+\Delta \varphi )
	\label{eq111}
\end{equation}

By Eq.\ref{eq111}, the nonlinear response of the Rydberg atom to the superposition field $E_{atoms}$ can be equivalent to envelop detection and after mixing quantum beat-note frequency signal $E_{IF}$ (Intermediate Frequency) containing the $E_{SIG}$ amplitude, frequency offset $\Delta f$  and phase offset $\Delta \varphi$. Once the coupling laser is set, the probe laser intensity will oscillate at a beat-note frequency, which means the weak $E_{SIG}$ signal can be detected. In this paper, only the $E_{\textup{atoms}}$ amplitude is discussed.

From Eq.\ref{eq44} and Eq.\ref{eq111} we have:
\begin{equation}
	\begin{aligned}[b]
		&     \Omega _{\textup{MW}}=\frac{\mu |E_{\textup{atoms}}|}{\hslash}=\frac{\mu (A+\textup{Bcos}(2\pi \Delta ft+\Delta \varphi))}{\hslash}\\
		&\quad      	=\frac{\mu A}{\hslash}+\frac{\mu \textup{Bcos}(2\pi \Delta ft+\Delta \varphi)}{\hslash}=\Omega _{\textup{DC}}+\Omega _{\textup{AC}}
	\end{aligned}	
	\label{eq5}
\end{equation}
where $\Omega _{\textup{DC}}$ is direct current(DC) component and $\Omega _{\textup{AC}}$ is alternating current(AC) component, respectively.

When the coupling laser is locked, as shown in Figure \ref{fig555}(a), the transmitted probe laser power $P_{\textup{T}}$ at the locked frequency point with different MW Rabi frequencies $\Omega _{\textup{MW}}$ is extracted. The relationship between the $P_{\textup{T}}$ and the $\Omega _{\textup{MW}}$ is obtained, as shown in the Figure \ref{fig555}(b). Additionally, the blue dashed line is the result of the red curve linear fitting in Figure \ref{fig555}(b), and the purple sine function curve at Figure \ref{fig555}(d) depicts the $\Omega _{\textup{MW}}$ with time t, and the orange sine function curve at Figure \ref{fig555}(c) depicts the $P_{\textup{T}}$ with time t. Figure \ref{fig555}(b) depicts that $P_{\textup{T}}$ decreases as the $\Omega _{\textup{MW}}$ increases, and $P_{\textup{T}}$ is linearly related to the $\Omega _{\textup{MW}}$ where the blue dashed line coincides with the red curve. When the $\Omega _{\textup{MW}}$ is too small or too large, influenced by the shape of the EIT-AT splitting spectrum, the $\Omega _{\textup{MW}}$ and the $P_{\textup{T}}$ will show an obvious non-linear relationship. When $\Omega _{\textup{MW}}$ is in the linear interval. The relationship between $P_{\textup{T}}$ and $\Omega _{\textup{DC}}$ and $\Omega _{\textup{AC}}$ can be expressed as: 
\begin{equation}
	P_{\textup{T}}=P_{\textup{TDC}}+P_{\textup{TAC}}=\beta \Omega _{\textup{DC}}+\kappa \Omega _{\textup{AC}}
	\label{eq6}
\end{equation}
where $\kappa(\beta)$ is the AC(DC) component ratio of the  $P_{\textup{TAC}}(P_{\textup{TDC}})$. $\kappa$ describes the variation rate of $P_{\textup{TAC}}$ with $\Omega _{\textup{AC}}$\cite{jing2020atomic}. Moreover, the $\kappa$ is adopted as an auxiliary reference index to characterize the superheterodyne sensor sensitivity changes. Based on Eq.\ref{eq6}, the greater value of $\kappa$, the stronger responsiveness of the Rydberg atom to the weak electric field, and thus the better sensitivity performance.

\begin{figure}[ht!]\centering
	\includegraphics[width=0.8\textwidth]{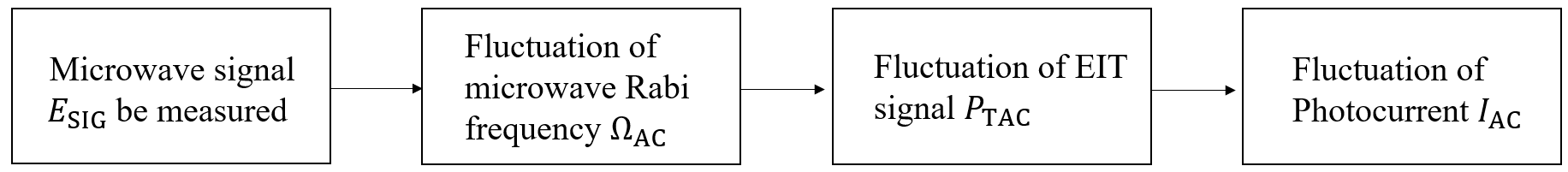}
	\caption{\label{fig6666}The process of receiving microwave signal by Rydberg atomic superheterodyne sensor}
\end{figure}

To summarize, the process of the weak $E_{\textup{SIG}}$ signal measured by the Rydberg atomic superheterodyne sensor can be represented in Figure \ref{fig6666}. The $E_{\textup{SIG}}$ is mapped by the Rydberg atom into fluctuations in the $\Omega _{\textup{AC}}$, which are further mapped into fluctuations in the $P_{\textup{TAC}}$ after locking the probe and coupling laser frequency, and then into fluctuations in the photocurrent signal $I _{\textup{AC}}$ via the photodetector. Finally, the signal in the form of the photocurrent signal power is output from the photodetector.

\begin{figure}[ht!]\centering
	\includegraphics[width=0.8\textwidth]{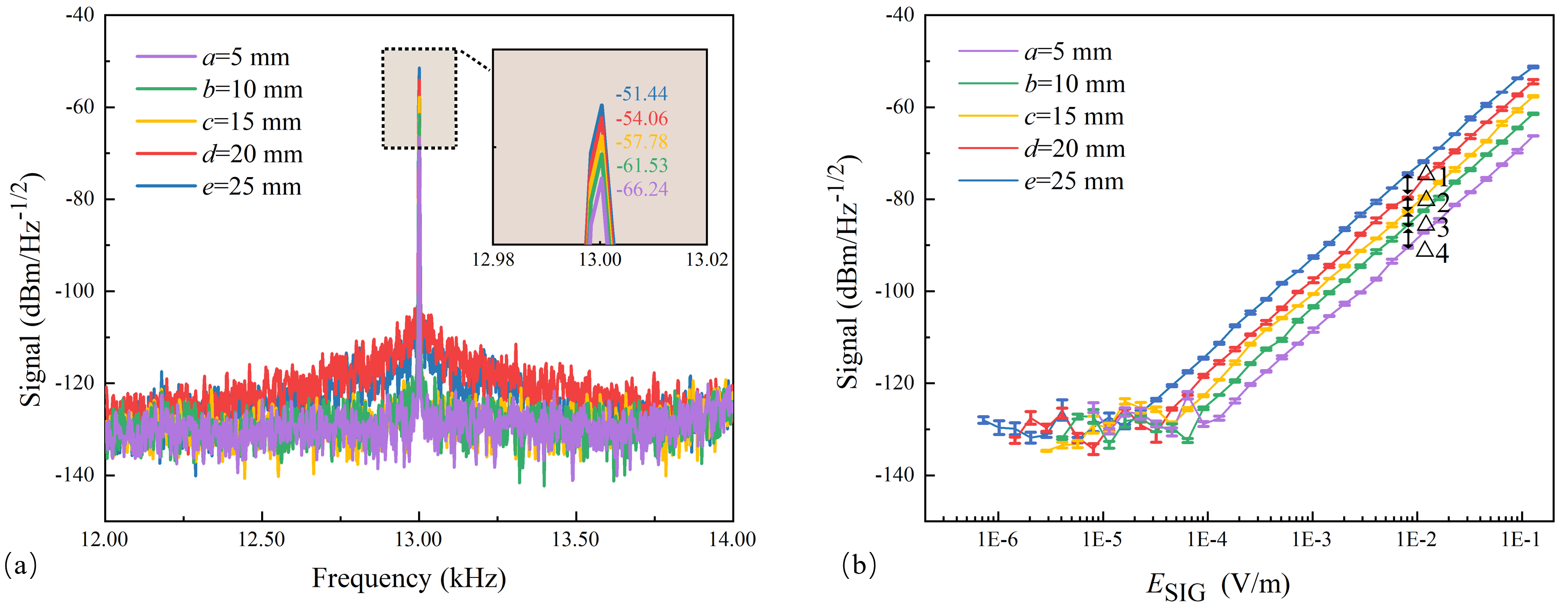}
	\caption{\label{fig7}(a)Measured spectrum analyzer results of 13 kHz detuning frequency signal. (b)Results from the Rydberg-atom superheterodyne, the output power plots of the spectrum analyzer as the $E_{SIG}$ functions, the error bars representing the standard deviation of sensitivity }
\end{figure}


Figure \ref{fig7}(a) depicts that the spectrum analyzer is utilized to gather the 13 kHz detuning frequency and amplitude of this beat-note signal by Rydberg-atom superheterodyne where purple curve, green curve, yellow curve, red curve, and blue curve refer to the 5 mm,10 mm,15 mm,20 mm,25 mm, respectively. Figure \ref{fig7}(b) illustrates that the beat-note signal intensity of the photodetector from the spectrum analyzer is approximately linear proportional to the $E_{SIG}$ for the cases with five different dimensions where $\Delta _{1}$, $ \Delta _{2}$, $ \Delta _{3}$, $ \Delta _{4}$ refer to the difference between blue curve and red curve($ 20 $ mm$\rightarrow$ $25$ mm), the difference between the red curve and yellow curve($ 15$ mm$\rightarrow$$20 $ mm ), the difference between the yellow curve and green curve($10$ mm$\rightarrow$ $15$ mm), the difference between the green curve and purple curve($5$ mm$\rightarrow$ $10$ mm), respectively. $\bar{\Delta}$ represents the mean value of the experimental data, where $\bar{\Delta_{1}}=4.1$ dB ,$\bar{\Delta_{2}}=3.2$ dB ,$\bar{\Delta_{3}}=2.9$ dB,$\bar{\Delta_{4}}=4.8$ dB. As the curves indicate, the beat-note signal power grows from a=5 mm to e=25 mm under the same laser power conditions. This reveals that atomic superheterodyne sensor sensitivity increases with the Rydberg atom population to a certain extent.
\begin{figure}[ht!]\centering
	\includegraphics[width=0.8\textwidth]{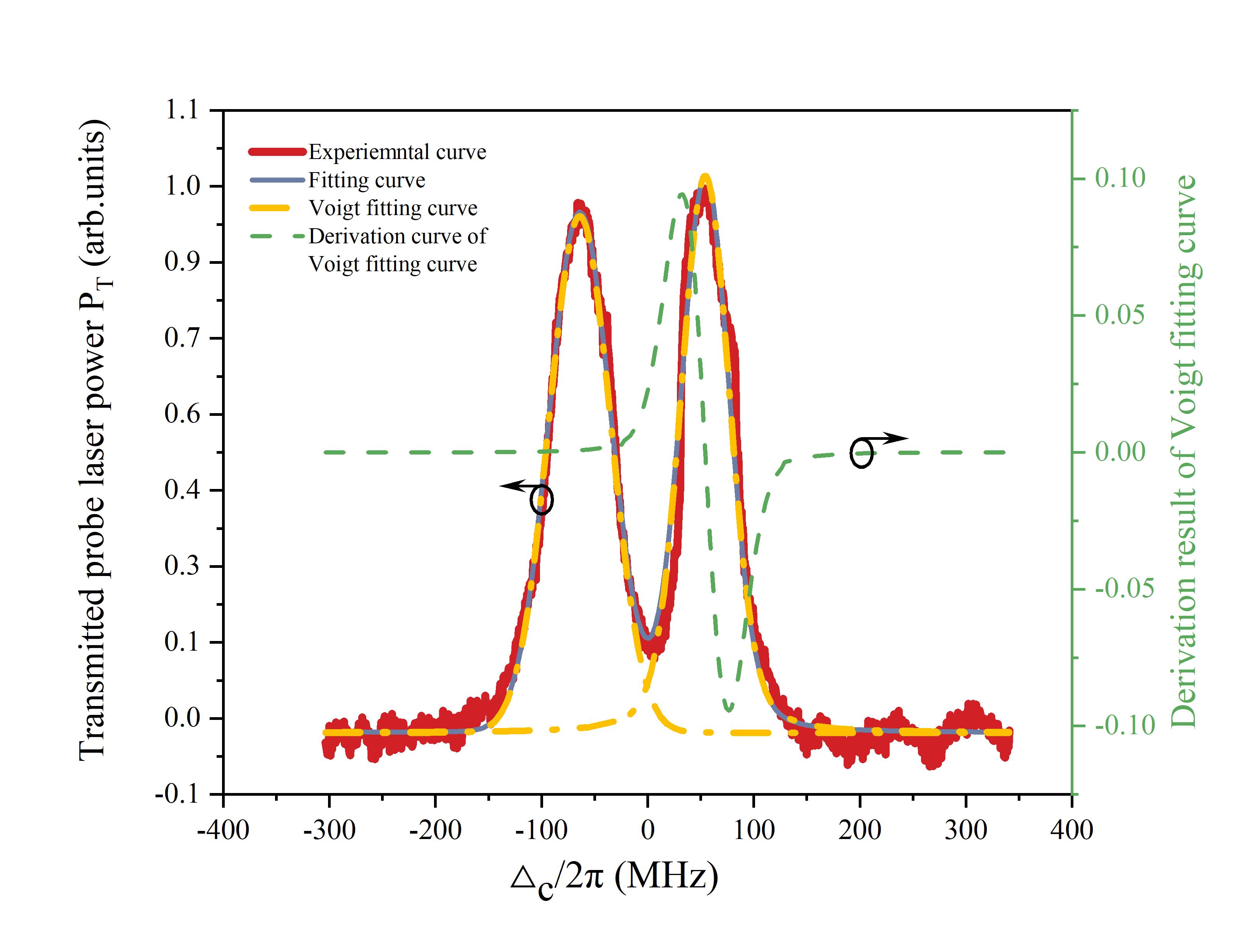}
	\caption{\label{fig8}Experimental EIT-AT signal versus coupling laser
		detuning. The cell length is 25 mm. Red curve presents experimental EIT-AT data. Yellow curve and blue curve present fitting curves. Green curve presents derivation curve of Voigt fitting curve }
\end{figure}

To thoroughly understand the reason of sensitivity experiences an upward trend from a=5 mm to e=25 mm, a series of E-field enhancement factors by $\kappa$ with five different dimensions are calculated based on the AT splitting approach, as shown in Figure \ref{fig8} and Table \ref{table2}. $\kappa$ can be calculated utilizing a Voigt function fitting to the experimental results. The yellow curves are two roughly identically Voigt function curves which are symmetrical regarding the $\Delta _{c}=0$. The blue curve is the fitting to the yellow curves, and the green curve is the derivative calculation of the Voigt function. Note that the orange and blue curves fit the EIT-AT spectrum well. Thus, to some extent, Voigt fitting curve can be utilized to describe the spectrum obtained experimentally. The frequency lock-in point value 0.024
($\Delta _{c}$=0) is the $\kappa$ in the green dashed line.

\begin{table}[!htbp]
	\centering 
	\caption{\label{table2}Performance comparisons of experimental(Exp) E-field enhancement factor and calculated(Cal) E-field enhancement factor by $\kappa$\  with the increment of cell length from 5 mm$\rightarrow$ 10 mm, 10 mm$\rightarrow$ 15 mm, 15 mm$\rightarrow$ 20 mm, 20 mm$\rightarrow$ 25 mm, respectively.}
	\begin{tabular}{ccc}
		\toprule  
		& Exp factor & Cal factor by $\kappa$\\ \midrule  
		$ 20$ mm$\rightarrow$ $25$ mm       & 1.6                           & 1.4                               \\
		$ 15$ mm$\rightarrow$ $20$ mm     & 1.4                            & 1.2                           \\
		$10$ mm$\rightarrow$ $15$ mm     & 1.4                            & 1.2                          \\
		$5$ mm$\rightarrow$ $10$ mm    & 1.7                            & 1.3                       \\ \bottomrule 
	\end{tabular}
\end{table}

Table \ref{table2} summarizes the specific performance comparison with experimental E-field enhancement factor($\sqrt{10^{\bar{\Delta}/10}}$) and calculated E-field enhancement factor by $\kappa$ with the increment of cell length. It is shown that the E-field enhancement factor is significantly improved with the increment of cell length, which equals superheterodyne sensor sensitivity growing with the increment of Rydberg atom population.
\section{Conclusion}
To summarize, we demonstrated that varying the path length under otherwise identical circumstances (notably Rydberg atom density and possibly the density of various background gases), is a strong experimental tool. Due to using a stepped  cell, consistency problem is solved for different cells. Importantly, we also revealed that the height of EIT, FWHM, and sensitivity grow with the increment of Rydberg atom population under the same laser power conditions to a certain extent.  It should be noted that despite the advantages of utilizing an cell as the atomic sensor, the impact of the cell itself on the measurement results is still the largest source of measurement uncertainty\cite{2015Effect}. Additionally, it is not the longer cell length, and the more the number of Rydberg atom population, the better the sensitivity. On the one hand, the longer length will require more probe and coupling laser power, leading to the destruction of the optimum power weak field approximation conditions and resulting in a poor integrated EIT; on the other hand, due to the Beer-Lambert law, the laser intensity  falls with the increasing transmission distance, resulting in significant difference between the probe and coupling laser power at the two ends of the cell. It leads to the descent of integrated EIT effect. Finally, due to the variation inside the vapor cell and absorption can ultimately enhance the overall uncertainty of a vapor-cell Rydberg-atom-based E-field measurement\cite{Haoquan}. But there is no denying that increasing the Rydberg atom population at relatively small cell length can improve weak E-field detection capability. We hope that the results can serve as an important step toward understanding the complex roles of Rydberg atom population effect on atomic sensors. 

\backmatter


\section*{Declarations}
\begin{itemize}
\item Funding: This work was supported by National Natural Science Foundation of China (Grant Nos. 12104509 and 62105338).
\item Conflict of interest/Competing interests: The authors declare no conflicts of interest.
\item Availability of data and materials: Data underlying the results presented in this paper are not publicly available at this time, but may be obtained from the authors upon reasonable request.

\end{itemize}



\bibliography{sn-bibliography}

\end{document}